\documentclass[
 aps,
]{revtex4-1}
\usepackage{amsmath}
\usepackage{amsfonts}
\usepackage{graphicx}
\usepackage{dcolumn}
\usepackage{bm}

\newtheorem{axiom}{Axiom}
\begin{document}
\title{The  Width Paradox  and the Internal Structure of a Black-Hole}
\author{Marcelo Schiffer}
 \affiliation{Physics Department,  Ariel University.}
 \email{schiffer@ariel.ac.il}
  \date{\today}

\begin{abstract}
In  the early days of Black Hole Thermodynamics, Bekenstein calculated  the mass dispersion of a macroscopic  black hole  that results from the stochasticity of the thermal radiation  it emits -- it  turned out to be negative for  black holes massive than $M \stackrel{>}{\sim} 10^{30}g$.   He  named it  the {\it  "mass width paradox"}. Here we revisit his early calculation, in an axiomatic approach with a set of more economical  assumptions and reach similar conclusions. We argue that the mass paradox results from considering a  black hole as a classical system, without an inner quantum structure.  As a matter of fact, when we take into account the discreteness of the area levels and assume    identical probability transition between  contiguous quantum states \cite{bekenstein}, the paradox disappears. In the process  we obtain the  probability of finding a black-hole  in some area eigenstate for a given averaged area.   As a by-product,   the  quantum scenario  also points towards a possible solution of the black hole information conundrum. Apparently we might have have killed two birds with a single stone. 
\begin{description}
\item[PACS numbers]:04.70.Bw, 04.70.Dy, 78.20.Ci, 78.45.+h
\end{description}
\end{abstract}

\pacs{Valid PACS appear here}
\maketitle

\section{The Width Paradox Revisited}
We start by defining  a set of very general postulates regarding the nature of black-hole mass fluctuations assuming the black hole is a classical system. 
\begin{axiom}:
For a macroscopic Schwarzschild black hole there is a smooth probability  distribution function $P(M,\overline{M})$. The first two moments are defined in the usual way,
\begin{subequations}
\begin{align}
\int_0^\infty P(M,\overline{M})dM=1 \label{normalization} \, ,\\
\overline{M}=\int_0^\infty M P(M,\overline{M})dM \label{average}\, ,\\
\overline{M^2}=\int_0^\infty M^2 P(M,\overline{M})dM\, . \label{average2}
\end{align}
\end{subequations} 
\end{axiom}
The dependence of the probability distribution for a Schwarzschild black hole on a single parameter  (the averaged mass) is dictated by the "no-hair theorem".
\begin{axiom}: For a macroscopic black hole the probability is sufficiently peaked around a very large mass such that:
\begin{eqnarray}
\lim_{M\to\infty} \frac{\partial^n P(M,\langle M \rangle)}{\partial M^n}=0 \, ,\\
\lim_{M\to\ 0} \frac{\partial^n P(M,\langle M \rangle)}{\partial M^n}=0\, ,
\end{eqnarray}
\label{boundary}
\end{axiom}
for all $n$. This condition allows us to disregard boundary terms when integrating by parts.
\begin{axiom}
The black hole mass dispersion is defined by
\begin{equation}
\Sigma^2(\overline{M}) =\overline{M^2} -\overline{M}^2\, ,
\end{equation}
\end{axiom}
where
\begin{equation}
\overline{M^2}=\int_0^\infty M^2 P(M,\overline{M})dM \, .
\end{equation} 
\begin{axiom}
The total energy (BH+radiation) is conserved in the radiation process. When a quanta in the energy mode $\epsilon_i$ is emitted the black hole mass decreases $M'=M-  \epsilon_i$ with $\epsilon_i<<M $
\end{axiom}

\begin{axiom}
The emission of the various modes are uncorrelated  and the emission probability in each mode is given by {\rm  (\cite{BekensteinMeisels})}
\begin{equation}
p(n_i)=(1-e^{-\gamma_i}) e^{-\gamma_i \epsilon_i}\, ,
\label{exponentialdistribution}
\end{equation}
where the mean number of quanta emitted in a specific mode satisfies {\rm \cite{donpage}}
\begin{equation}
\langle n_i \rangle =\frac{\Gamma(\overline{M},\epsilon_i)}{e^{8\pi G \overline{M}\epsilon_i/\hbar}-1}\, .
\label{averagequanta}
\end{equation}
{\rm Here, $\overline{M} $ represents the averaged black hole mass and  $\Gamma(\overline{M} \epsilon_i)$ is the black hole absorptivity for a given  mode.}
\end{axiom}
\begin{axiom}
Laking  any internal degrees of freedom, black hole  mass  fluctuations  can result exclusively  from the stochastic nature of the Hawking radiation. 
\end{axiom}
Please notice: we reserve the overline for average symbol  obtained via $P(M, \overline{M})$ (black hole average) while the $\langle\rangle$ is used for averages obtained via $p(n_i)$ (radiation average).
\section{The mass dispersion}

A black hole whose original mass is  $M^*$ emits quanta in various energy modes . Its mass decreases by $M=M^*-\sum_i n_i \epsilon_i$, where $n_i$ is the the number of quanta emitted in the $\epsilon_i$ mode. The other way around, the original mass is $M^*=M+\sum_i n_i \epsilon_i$. Thus, after de emission 
\begin{equation}
P(M,\overline{M})= \sum_{\{n_i\}} P(M+\sum_i \epsilon_i n_i,\overline{M}^* )\prod_i p_i(\epsilon_i) \, .
\end{equation}
where postulates (4),(5),(6) were called for. We  expand the argument of the probability distribution around $M$, 
\begin{eqnarray}
P(M,\overline{M})&=&P(M,\overline{M^*})+\ P'(M,\overline{M^*}) \sum_i  \sum_{n_i} \epsilon_i n_i p(n_i)\\
&+&\frac{1}{2} P'' (M,\overline{M^T*}) \left[\sum_i  \sum_{n_i} \epsilon_i ^2 n_i^2 p(n_i)+\sum_{i\neq j} \epsilon_i \epsilon_j  \sum_{n_i,n_j} n_i n_j p(n_i)p(n_j)\right] +\sum_{N>2}\frac{\Delta_N}{N!}P^{(N)}(M,M^*)\, .
\end{eqnarray}
Primes denote first and second derivatives of $P(M,\overline{M})$ with respect to the {\it{ actual mass} $M$},  $P^{(N)}$ represents  higher order derivatives and $\Delta_N$  higher moments .
Since
\begin{eqnarray}
 \sum_i  \sum_{n_i} \epsilon_i n_i p(n_i) &= &\sum_i  \epsilon_i \langle n _i \rangle \, ,  \\
  \sum_i  \sum_{n_i} \epsilon_i ^2n_i^2 p(n_i) &= &\sum_i  \epsilon_i ^2\langle n_i ^2\rangle    \,
 \end{eqnarray}
 and furthermore
  \begin{eqnarray}
\sum_{i\neq j}  \sum_{n_i ,n_j}\epsilon_i \epsilon_j n_i n_j p(n_i)p(n_j) =\sum_{i\neq j}\epsilon_i\epsilon_j \langle n\rangle _i \langle n_j
\rangle =\sum_i \epsilon_i \langle n_i \rangle \left( \sum_j\epsilon_j \langle n\rangle _j - \epsilon_i \langle n _i\rangle \right) \, ,
\end{eqnarray}
It follows that
\begin{eqnarray}
P(M,\overline{M})-P(M,\langle M^*\rangle)=P'(M,\overline{M^*})\langle  \epsilon \rangle 
+\frac{1}{2} \left[\sigma^2 +
\langle  \epsilon\rangle ^2 \right] P''(M,\overline{M}^*\rangle) +\sum_{N>2}\frac{\Delta_N}{N!}P^{(N)}(M,M^*) \, ,
 \label{difference}
\end{eqnarray}
where   we defined the mean energy of the radiation  emitted in all modes and the corresponding deviation
\begin{eqnarray}
\langle \epsilon \rangle  \equiv \sum_i \epsilon_i \langle n_i\rangle \\ 
\sigma^2\equiv \sum_i \epsilon_i ^2 \left(\langle n_i^2\rangle  -\langle n_i^2\rangle \right) \, .
\end{eqnarray}
All results in this section stems from the above equation. Indeed, integrating both sides of eq.(\ref{difference}) and recalling eq.(\ref{normalization}):
\begin{equation}
\int_0^\infty P'(M,\overline{M}^*) \epsilon  dM
+\frac{1}{2}\int_0^\infty   P''(M,\overline{M}^*) \left[\sigma^2 +
 \epsilon^2 \right] dM+\frac{1}{N!}  \sum_{N>2} \int_0^\infty P^{(N)}(M,\overline{M}^*) \Delta_N dM=0 \, .
\end{equation}
 This expression should vanish identically.  This happens iff    $\langle\epsilon\rangle , \sigma^2$ and$  \Delta_N$  can be moved outside the integrals ( all the derivatives vanish at infinity) . Consequenlty  all the moments must depend upon the (black hole)  averaged mass $\overline{M}$  but not upon its  instantaneous mass $M$. 
Let us multiplicate eq.[\ref{difference}] by the mass $M$ and integrate it:
\begin{eqnarray}
\Delta \overline{M} \equiv \overline{M}^* - \overline{M} = - \langle \epsilon \rangle   \int_0^\infty M  P'(M,\langle M^*\rangle) dM \nonumber \\
-\frac{1}{2} \left[\sigma^2 +
\langle \epsilon\rangle^2\right]  \int_0^\infty M  P''(M,\overline{M}^*\rangle)dM
  -  \sum_{N>2}\frac{\Delta_N}{N!}  \int_0^\infty M  P^{(N)}(M,\overline{M}^*) \, .
 \end{eqnarray}
Integrating once these integrals by parts, only the first one remains.  The change of the black hole average mass is identical to the average energy of the emitted radiation, as as expected from energy conservation grounds:
\begin{equation}
 \overline{\Delta M}= \langle \epsilon \rangle \, .
\label{energyconserv}
\end{equation}
Last, we calculate the bh mass dispersion 
\begin{equation}
\Sigma^2(\overline{M})=\overline{M^2}-\overline{M}^2\, .
\end{equation}
For this end, we multiply eq.(\ref{difference}) by $M^2$ and integrate over the mass, 
  \begin{eqnarray}
 \overline{M^{*2}} - \overline{M^{2}}  &=&- \langle \epsilon\rangle    \int_0^\infty M^2  P'(M,\overline{M}^*) dM\nonumber\\
-\frac{1}{2} \left[\sigma^2 +
\langle  \epsilon\rangle ^2\right]  \int_0^\infty M^2  P''(M,\overline{M}^*)dM
  &-& \sum_{N>2}\frac{\Delta_N  }{N!}\int_0^\infty M^2  P^{(N)}(M,\overline{M})=0 \, .
 \end{eqnarray}
Integrating  twice this expression by parts, the last integral drops out and we are left with
 \begin{equation}
\overline{M}^{*2} -\overline{M^{2}} =2 \overline{M}  \langle \epsilon \rangle 
- \sigma^2 - \langle \epsilon\rangle ^2\, .
 \end{equation}
Together with  the identity:
\begin{equation}
\overline{M^{*2}} -\overline{M^{2}}= \Sigma^{*2}-\Sigma^{2}+2\langle M\rangle \epsilon +\epsilon^2 \, ,
\label{identity}
\end{equation}
where  $\Sigma^{*2}=\Sigma^2(\overline{M}^*)$,  gives:
\begin{equation}
\Sigma^{*2}-\Sigma^{2}=-\sigma^2 -2 \epsilon^2 \, .
\end{equation}
Clearly for a macroscopic black hole black hole $ \epsilon <<\langle M \rangle$ ,  
\begin{equation}
\Sigma^{*2}-\Sigma^{2}\approx \frac{\partial \Sigma^{2}}{\partial \langle M \rangle } \epsilon
\end{equation}
So,
\begin{equation}
 \frac{\partial \Sigma^{2}}{\partial \langle M \rangle } =-\left( \frac{\sigma^2}{\epsilon} +2\epsilon  \right) \, .
\label{mysigma}
\end{equation}
This equation relates the  dispersion of the black hole to statistical properties of the emitted radiation (average energy and dispersion).
We move on by calculating the rhs of the previous equation for the emitted radiation.  As we mentioned  earlier the parameter $\gamma_i$ in the distribution eq.[\ref{exponentialdistribution}]  is obtained by matching the mean number of emitted quanta
\begin{equation}
\overline{n}_i=\frac{1}{e^{\gamma_i}-1}=\frac{\Gamma_i}{e^{x_i }-1} \, ,
\end{equation}
where $x_i=8\pi \overline{M} \epsilon_i/m_p^2$. Clearly, 
\begin{equation}
\langle \epsilon\rangle =\sum_j \frac{\epsilon_i \Gamma_i}{e^{x_i }-1}=\left(\frac{m_p^2}{8\pi \overline{M}}\right) \sum_j \frac{x_i \Gamma_i}{e^{x_i }-1}
\end{equation}
and
\begin{equation}
\sigma^2=\sum \epsilon_i^2\left(\langle n_i^2\rangle -\langle n_i \rangle^2\right)=\sum \epsilon_i^2\frac{e^{\gamma_i}}{(e^{\gamma_i}-1)^2}=\left(\frac{m_p^2}{8\pi \overline{M}}\right)^2 \sum  
\left(\frac{x_i^2}{(e^{\gamma_i}-1)^2}+\frac{x_i^2}{e^{\gamma_i}-1}\right)\, .
\end{equation}
Equivalently, 
\begin{equation}
\sigma^2=\left(\frac{m_p^2}{8\pi \overline{M}}\right)^2 \sum  
\left(\frac{x_i^2 \Gamma_i^2}{(e^{x_i}-1)^2}+\frac{x_i^2\Gamma_i}{e^{x_i}-1}\right)\, .
\end{equation}
For macroscopic black holes whose mass $M >>m_e/m_p^2 \sim 10^{17}g$ ($m_e$ stands for the electron mass) only  massless quanta are emitted as the emission of massive particles is exponentially suppressed.  Lacking any additional dimensional parameter, the black hole absorption coefficient $\Gamma_i=\Gamma(\overline{M}\epsilon_i/m_p^2)=\Gamma(x_i)$. Furthermore, it is safe to take the  continuous approximation  and translate sums into integrals:
\begin{equation}
\langle \epsilon\rangle =A \left(\frac{m_p^2}{8\pi \overline{M}}\right)   \quad,\quad \sigma^2 = B \left(\frac{m_p^2}{8\pi \overline{M}}\right)^2 \, ,
\end{equation}
where $A$ and $B$ are numerical coefficients:
\begin{equation}
A=\int_0^\infty  \frac{x \Gamma(x)}{e^{x }-1}dx \quad ; \quad B=
\int_0^\infty x^2 \Gamma(x) \frac{\Gamma(x)+e^x-1}{(e^{x}-1)^2}dx \, ,
\end{equation}
\begin{equation}
\Sigma^2 = \Sigma_0^2-\left(\frac{m_p^2}{8\pi}\right)\left(\frac{B}{A}+2A\right)\ln\frac{\overline{M}}{\overline{M}_0}\, /
\end{equation}
Our result reproduces to Bekenstein's original calculation, with different numerical coefficients.
In the above expression $\Sigma^2$ and $\overline{M}_0$ are the mass dispersion and average mass at some reference point where the black hole is emitting massless quanta alone ($\overline{M}_0 > 10^{17}g$) and  $A$ and $B$ are dimensionless parameters of order one. At this reference point only massless quanta are being emitted, in the lack of additional   parameters with dimensions of mass besides the Planck mass   necessariely 
$\Sigma_0^2 \sim m_p^2$: for sufficiently massive black holes the mass dispersion becomes eventually  negative.  
This is the width paradox. In the original paper it was argued that the paradox can be circumvented 
if the average number of emitted quanta is a function of the actual mass $M$  instead of the averaged mass $\langle M\rangle$ . But, this as we have seen this is at odds with the normalization condition and energy conservation ($\Delta \overline{M}= \langle \epsilon\rangle $).

\section{The black hole has an inner structure}

The above  paradoxical result arises   from  the assumption  the black hole is a structureless  object that emits grey-body radiation at the Bekenstein-Hawking temperature. The situation reminds the early days of quantum theory, where Planck derived the black body radiation from the absorption and emission of radiation by harmonic oscillators that undergo  transitions between discrete energy levels.  Instead of harmonic oscillators, following Bekenstein-Mukhanov's proposal \cite{mukhanov} we assume that the black hole area is quantized in units of the Planck length.  Evidence of area quantization arises from various different  theoretical approaches like   loop quantum gravity,\cite{loop}  or more qualitative arguments like the gedanken experiment of  absorption of quanta \cite{gedanken}, and also excitation of the black hole's ring modes\cite{ringmodes}. Recently it has been claimed that the late echo detected in the merging of black holes might be an imprint of the area quantization \cite{gravecho}. Any way:
\begin{equation}
A(n)=4 \kappa l_p^2 n \, .
\label{qarea}
\end{equation}
Here  $\kappa$ is a numerical parameter of order one. In the lack of a comprehensive theory of quantum gravity, its value depends upon the quantization argument. Very much like in atomic physics, emission of quanta results from transitions among energy levels. The  mass, angular momentum and entropy changes due to the emission  of a quantum in the continuous approximation satisfy the first law:
\begin{equation}
\delta M= T_{bh} \Delta S+\Omega_{bh} \delta J \, ,
\end{equation}
where $\delta M=-\hbar \omega$ and $\delta J=-m \hbar \quad (m=0,\pm1,\pm 2...)$ are the energy and angular momentum of  the emitted  quantum. Recalling that $A_{bh}=4 l_p^2 S_{bh}$, in the transition among contiguous  area levels the quantum numbers of the emitted photon emission satisfy the constraint
\begin{equation}
x \equiv \frac{\hbar (\omega - m \Omega_{bh})}{T_{bh} }= \kappa \, .
\end{equation}
It is  assumed \cite{bekenstein} that further transitions occur in a cascade chain $n-1\rightarrow n-2;n-2\rightarrow n-3,\dots$. For a macroscopic  black hole $T_{bh}$ remains practically constant along this chain, meaning that the radiation is very nearly monochromatic. Accordingly, as  the black hole cascades into lower levels, say from the $n\to m$  it emits radiation consisting of effectively monochromatic  $n-m$ quanta. The very nearly monochromatic feature of the radiation means that  the timescale for the emission is very large and  adiabaticity   ensures that the transition probability remains very nearly unchanged in the following transitions.  Call  $e^{-\alpha}$ the black hole decay probability to a contiguous level. Accordingly, the transition probability from $n$ to $m$ levels is
\begin{equation}
W_{n \to m} = C e^{-\alpha (n-m)} \qquad n \geq m \, .
\end{equation}
 The normalization of the transition probability
 \begin{equation}
 \sum_{m=0}^n W_{n \to m} =1 
 \end{equation}
requires $C=\frac{1-e^{-\alpha}}{1-e^{-(n+1)\alpha}}$;  nevertheless in the limit  $n>>1$ the normalization constant $C \approx 1-e^{-\alpha}$ , the system is effectively  translational invariant.  In other words, $W_{n\rightarrow m}$ represents the probability of a black hole decaying from $n\rightarrow m$ level and    $1-e^{-\alpha}$ represents  the probability of not making any additional transition and  remaining in that state. Let $P_n(t) $ represent the probability of finding the black hole in a specific area level at time $t$  measured in some unit of time. Then, at time $t+1$ after  undergoing a  transition :
 \begin{equation}
 P_m(t+1)=\sum_{n\geq m} W_{n\to m} P_n(t) \, .
 \label{recursion}
 \end{equation} 
From this we can calculate the area change after each emission
\begin{eqnarray}
\langle m(t+1)\rangle &=&\sum_m m P_m(t+1)=\sum_{n\geq m} m W_{n\to m} P_n(t)=(1-e^{-\alpha}) \sum_n P_n(t) e^{-\alpha n}\sum_{m\leq n} m e^{\alpha m}\\
&=&(1-e^{-\alpha}) \sum_n P_n(t) e^{-\alpha n}\frac{d}{d\alpha}  \frac{1-e^{\alpha (n+1)}}{1-e^\alpha}= \sum_n P_n(t) \left[n-\frac{1}{e^\alpha-1}+\frac{e^{-\alpha n }}{e^\alpha-1)} \right] \, .
\end{eqnarray}
Now, the probability distribution for a macroscopic black hole peaks for some very large value of $n$ in which case the  exponential term is largely suppressed . 
Thus,  the average area   change in each step is constant
\begin{equation}
\langle A(t+1)\rangle= \langle A(t) \rangle-\frac{4kl_p^2}{e^\alpha-1} \, .
\label{areastep}
\end{equation}
Starting from a given initial reference time $t=0$, the area decreases linearly:
\begin{equation}
\langle A(t+1)\rangle= \langle A(0) \rangle- t \frac{4kl_p^2}{e^\alpha-1} \, .
\end{equation}
From similar considerations regarding
\begin{eqnarray}
\langle m^2(t+1)\rangle& =&(1-e^{-\alpha}) \sum_n P_n(t) e^{-\alpha n}\frac{d^2}{d\alpha^2}  \frac{1-e^{\alpha (n+1)}}{1-e^\alpha}\\
&=& \sum_n P_n(t) \left[n^2-\frac{2 n}{e^\alpha-1}-e^{-n \alpha} \frac{e^\alpha+1}{(e^\alpha-1)^2}+\frac{e^\alpha+1}{(e^\alpha-1)^2} \right] \, ,
\end{eqnarray}
or 
\begin{equation}
\langle m^2 (t+1)\rangle  = \langle m^2 (t)\rangle  - \frac{2 \langle m(t)\rangle}{e^\alpha -1}  + \frac{e^\alpha+1}{(e^\alpha-1)^2} \, ,
\end{equation}
and we dropped  again the exponential term. The area dispersion grows linearly with time
\begin{equation}
\Delta A^2 (t+1) =\Delta A^2(0)+\frac{16 k^2 l_p^4}{1-e^{-\alpha}}t \, ,
\end{equation}
very much like in the one dimensional random walk. Can we obtain the probability distribution of the various black hole area levels ? Let us assume that at the initial time $t=0$ (say, at the horizon formation ) the black hole in an area eigenstate $n_0$, that is to say, $P_n(0)=\delta_{n,n_0}$. The first iteration of eq.(\ref{recursion}) gives
\begin{equation}
 P_m(1)=(1-e^{-\alpha}) e^{-\alpha (n_0-m)} \, .
\end{equation}
It can be easily checked that further iterations preserve this general form
\begin{equation}
P_m(t)= A_m(t)(1-e^{-\alpha})^t e^{-\alpha (n_0-m)} \, ,
\end{equation}
where 
\begin{equation}
A_m(t+1)=\sum_{n<m} A_n(t) \, .
\end{equation}
The easiest way to find $A_m$ is through the normalisation condition $\sum_m P_m(t)=1$. Defining $z=e^{-\alpha}$ and  reindexing $s=n_0-m$ it follows that
\begin{equation}
\sum_{s\geq 0} A_{n_0-s}(t)z^s=\frac{1}{(1-z)^t} \, .
\end{equation}
Consequently
\begin{equation}
A_{n_0-s}(t)=\frac{1}{s!} \left.\left(\frac{d}{dz}\right)^s (1-z)^{-t}\right|_{z=0}={t+s-1\choose{t-1}  } \, ,
\end{equation}
and the probability of finding the black hole  in a given area eigenstate $m$ is negative binomial:
\begin{equation}
P_m(t)= {t+n_0-m-1\choose{n_0-m} }(1-e^{-\alpha})^t e^{-\alpha (n_0-m)} \, .
\label{nbinomial}
\end{equation}
Here it is tacitly assumed that $t <<n_0$, as to avoid approaching the  Planckian region . The black hole mass deviation  entails the calculation of $\langle m \rangle -\langle \sqrt{m}\rangle^2 $.  The inverse Laplace transform provides an integral representation of $\sqrt{m}$ :
\begin{equation}
\sqrt{m}=\frac{1}{2\pi i}\int_{\gamma -i \infty}^{\gamma +i \infty} \frac{\sqrt{\pi}}{2}\frac{dz}{z^{3/2}} e^{m z} \, ,
\end{equation}
where the integration runs at a vertical axis in the complex plane at distance $\gamma$ from the origin such that  all poles  remain to the left side of this line. Thus,
\begin{equation}
\langle \sqrt{m(t)}\rangle= \sum_{m\leq n_0} \sqrt{m}P_m(t)=- \frac{i}{4\sqrt{\pi}}(1-e^{-\alpha})^t \int_{\gamma -i \infty}^{\gamma +i \infty} \frac{dz}{z^{3/2}} \sum_{s=0}^\infty e^{(n_0-s) z}{t+s-1\choose{s} }e^{-\alpha s } \, ,
\end{equation}
which can be rewritten as
\begin{equation}
\langle \sqrt{m(t)}\rangle=- \frac{i}{4\sqrt{\pi}} (1-e^{-\alpha})^t  \int_{\gamma -i \infty}^{\gamma +i \infty} \frac{dz}{z^{3/2}} e^{n_0 z}\frac{1}{\left(1-e^{-(\alpha+z) }\right)^t} \sum_{s=0}^\infty  {t+s-1\choose{s} }\left(1-e^{-(\alpha+z) }\right)^t e^{-(\alpha+z) s } \, .
\end{equation}
The  sum  is formally the normalization of the binomial probability distribution, therefore
\begin{equation}
\langle \sqrt{m(t)}\rangle=- \frac{i}{4\sqrt{\pi}} (1-e^{-\alpha})^t I
\end{equation}
with
\begin{equation}
I=\int_{\gamma -i \infty}^{\gamma +i \infty} \frac{e^{n_0 z}}{z^{3/2} (1-e^{-(z+\alpha) })^t} d z\, .
\end{equation}
The poles of this expression are located at $z=-\alpha + 2\pi n i,\alpha >0 $ and accordingly the integration is to be performed along  any vertical line with $\Im(z)= \gamma > 0$. Notice that there is a branch cut running from  $z\rightarrow -\infty$ to $z=0$, $\mathbb{R}^-$. 
\begin{equation}
\int_{\Gamma} \frac{e^{n_0 z}}{z^{3/2} (1-e^{-(z+\alpha) })^t} d z=2\pi i \sum Residues \, .
\end{equation}
The contour integral  is $\Gamma=\gamma+ C_R +L_1+L_2+C_r$ with $R\rightarrow \infty$; $L_1$ runs from $-\infty$ to the origin above the branch cut and  $L_2$ runs back to $-\infty$ beneath the branch cut and $C_r$ is the semi-circle connecting these lines with $r\rightarrow 0$.  Last, $C_R \rightarrow 0$ is the semi circle with
 $R \rightarrow \infty$.  According to Jordan's lemma  $C_R \rightarrow 0$. In the principal branch $z=|z|  e^{i\theta}, -\pi <\theta< \pi$. For convenience we define 
 \begin{equation}
 h(z)=\frac{e^{n_0 z}}{(1-e^{-(z+\alpha) })^t}\, .
 \end{equation}
 After integration by parts:
\begin{equation}
\int_{L_1+L_2+C_r}\frac{dz}{z^{3/2} } h(z)=-2\left. \frac{h(z)}{\sqrt{z}}\right|_{-\infty+i\pi }^{-\infty-i \pi} +2\int_{-\infty}^{0}\frac{dx}{(-x)^{1/2} (e^{i \pi/2}-e^{-i\pi/2})} h'(x)+2\int_{C_r}\frac{dz}{(z)^{1/2} } h'(z)\, .
\end{equation} 
The first term vanishes and so does the last one for an infinitesimal semi-circle around the origin. Changing $x\rightarrow -y^2$
\begin{equation}
\int_{L_1+L_2+C_r}\frac{dz}{z^{3/2} } h(z)=-i \int_{-\infty}^\infty dy h'(-y^2)\, .
\end{equation}
Explicitly,
\begin{equation}
h'(-y^2)=\left(\frac{\sqrt{n_0 \pi}}{(1-e^{y^2-\alpha})^t} -t \sqrt{\frac{\pi}{n_0}} \frac{e^{y^2-\alpha}}{(1-e^{y^2-\alpha})^{t+1}}\right)\delta(y)\, ,
\end{equation}
and we  defined the function
\begin{equation}
\delta(y)=\sqrt{\frac{n_0}{\pi}}e^{-n_0 y^2}\, .
\end{equation}
Now, $n_0\sim A/l_p^2$ is a huge number,  for all purposes the above function is the Gaussian representation of  the delta function. Thus:
\begin{equation}
\int_{L_1+L_2+C_r}\frac{dz}{z^{3/2} } h(z)=i \left(\frac{\sqrt{n_0 \pi}}{(1-e^{-\alpha})^t} -t \sqrt{\frac{\pi}{n_0}} \frac{e^{-\alpha}}{(1-e^{-\alpha})^{t+1}}\right)\, .
\end{equation}
The poles of $h(z)$ at  $z=-\alpha + 2n\pi i$  are  of order $t$ and accordingly  the residues are:
\begin{equation}
\frac{1}{(t-1)!} \frac{d^{t-1}}{dz^{t-1} }  \left.\frac{e^{n_0 z}}{z^{3/2}}\right|_{-\alpha + 2n \pi i}\, .
\end{equation}
This expression is proportional to $e^{-n_0 \alpha} $, the contribution of the residues  to the integral is exponentially small. Putting all these pieces together
\begin{equation}
\langle \sqrt{m(t)}\rangle =\sqrt{n_0}-\frac{t}{\sqrt{n_0}} \frac{1}{e^\alpha-1} \, .
\end{equation}
From eq.(\ref{areastep}) we know that
\begin{equation}
 \langle m \rangle =n_0 -\frac{t}{e^\alpha-1}\, ,
 \end{equation}
therefore,
 \begin{equation}
 \Delta M= \sqrt{\frac{k}{4\pi} } m_p \left(\frac{t}{e^\alpha -1} -\frac{t^2}{A_0}  \frac{4\kappa l_p^2}{(e^\alpha-1)^2}\right)\, .
 \end{equation}
 This expression   never becomes negative as we are assuming  that  $t<<n_0$.   What is the time scale for each one of these transitions ? The black hole emissivity  $\sim M^{-4}$ multiplied by the horizon  area gives a mass loss rate  $\dot{M}\sim -M^{-2} $ or equivalently the area loss rate 
 $\dot{A}\sim -M^{-1} $ . Calling the time scale of each transition $\tau$, comparisson to eq.(\ref{areastep}) gives the estimative 
 \begin{equation}
 \tau \sim  \frac{\langle M\rangle  G}{ e^\alpha -1}
 \end{equation}

As a last remark, the value of the one step decay probability $e^{-\alpha}$ can be obtained  relying on the correspondence principle. The average number of transitions  must match the mean number of quanta emitted in the corresponding mode
 \begin{equation}
 \frac{1}{e^\alpha -1}=\frac{\Gamma(x)}{e^x-1}\, .
 \end{equation}

\section{Eternal Black Holes}

The probability  $P_m(t)$ obtained in eq(\ref{nbinomial})  for an initial eigenstate $n_0$ is  the conditional  probability of finding  the black hole in an area eigenstate $m$ at time $t$ , given that it started from to some  initial  state $n_0$. Replacing $n_0 \rightarrow n$ it represents the conditional probability $P_t (m|n)$ of finding  at time $t$ the black hole at state $m$ given that it was initially at the state $n$.  For an eternal black hole,   there is no initial eigenstate , any moment can be rearded as the initial time. Put in another way, the probability distribution of finding the black hole in a given eigenstate must be  a universal function, calculated at different times.  Let us call the probability distribution at time $t$ as $q_m(t)$
Clearly:
\begin{equation}
q_m(t+\tau)=\sum_{m\leq n} P_\tau(m|n) q_n((t)\, .
\end{equation}
More explicitly 
\begin{equation}
q_m(t+\tau)= \sum_{n\leq m}{\tau +n-m-1\choose{n-m} }(1-e^{-\alpha})^\tau e^{-\alpha (n-m)} q_n(t) \, .
\end{equation}
Multiplying this equation by $e^{i m\varphi} $, summing both sides over $m$ while defining the generating function $ Q(t,\varphi)= \sum q_m(t) e^{i m\varphi} $ gives:
\begin{equation}
Q(t+\tau,\varphi )=  \sum_m \sum_{n\leq m}{\tau +n-m-1\choose{n -m} }(1-e^{-\alpha})^\tau e^{-\alpha (n-m)}  e^{i m\varphi}q_n(t)\, .
\end{equation}
Defining $s=n-m$, after reshuffling indexes :
\begin{equation}
Q(t+\tau,\varphi)= (1-e^{-\alpha})^\tau \left[  \sum_{s\geq0}{\tau + s-1\choose{s} } \left(e^{-\alpha}  e^{-i\varphi }\right)^s \right] Q(t, \varphi)\, .
\end{equation}
With the aid of the identity
\begin{equation}
\sum_{r\geq 0} {\tau + s-1\choose{s} }x^r=\frac{1}{(1-x)^\tau}\, ,
\end{equation}
we can write
\begin{equation}
Q(t+\tau,\varphi)= \left(\frac{1-e^{-\alpha}}{1-e^{-i  \varphi } e^{-\alpha}}\right)^\tau  Q(t,\varphi) \, .
\end{equation}
Normalization of the  probabilities calls for $Q(t,0)=1$  :
\begin{equation}
Q(t,\varphi)=\left( \frac{1-e^{-\alpha}}{1- e^{-\alpha}e^{-i \varphi} }\right)^t \, .
\end{equation}
The probability distribution are the Fourier coefficients of the generating function
\begin{equation}
q_m(t) =\left(1-e^{-\alpha}\right)^t\frac{1}{2\pi}\int_0^{2\pi} \frac{e^{i m \varphi}}{(1- e^{-\alpha}e^{-i \varphi} )^t }d\varphi  \, .
\end{equation}
Calling $z=e^{i \varphi}$
\begin{equation}
q_m(t) =-\left(1-e^{-\alpha}\right)^t\frac{i}{2\pi i}\oint  \frac{z^{t+m-1}}{(z- e^{-\alpha} )^t } dz \, .
\end{equation}
From the residues theorem it follows that the distribution is also negative binomial
\begin{equation}
q_m(t)={t +m-1\choose{t--1}}\left(1-e^{-\alpha}\right)^{t }e^{-\alpha m}  \, .
\label{eternal}
\end{equation}
Clearly the time  $t$ is not a physically sound variable, but we can parametrize the evolution as a function of the average area.

\section{The Information Paradox  and Concluding Remarks}
Assuming that (i) the black hole radiation is grey body;  that (ii)  its  mass fluctuations originate from the randomness   of the radiation emitted and  (iii) that the black hole has no inner structure  leads to the paradoxical situation were mass fluctuations become negative for sufficiently large black holes. The discrepancy  is solved assuming that the black hole has internal degrees of freedom ( the various discrete  quantum area eigenstates ) and an a  constant transition probability among contiguous area levels.  From a different perspective, should a black hole reach  thermal equilibrium with its own radiation  the detailed balance condition must hold and this also requires discrete energy  levels. Either way, consistency with Quantum Statistics demands discretization of black hole area states. The  probability distribution we obtained for an eternal black hole  was based on very solid premisses and  a robust calculation.This is a true quantum gravity result that should be substantiated  someday by a full quantum theory of gravity .

For an eternal black hole  the probability distribution should not be calculated at time  $t$, which clearly  is not a useful parameter as the origin of time can always be shifted; the averaged area is the only meaningful parameter to express the probability distribution.  Shannon's information of  an eternal  black hole should be regarded as pure noise, it has no truly informational content. 

Naturally, the information content of a black hole formed at some time
 must be represented by subtracting from  its  Shannon's information the noise, that is to say, Shannon's information of the eternal black hole (both compared for the same value of averaged horizon area). Specifically 
\begin{equation}
I(t)=I_{BH}(t)  - I_{EBH} \, ,
\end{equation}
where
\begin{equation}
I_{BH}(t)=-\sum_n \Pi_n(t) \ln \Pi_n(t) \, ,
\end{equation}
with
\begin{equation}
\Pi_n(t)= - \sum_m p_t(n|m) \Pi_m(0) \, ,
\label{convolution}
\end{equation}
where $\Pi_m(0)$ represents the quantum state (the probability distribution )at the time the horizon is formed. Furthermore
\begin{equation}
I_{EBH }=  - \sum_m q_n(t')  \ln q_m(t')  \, .
\end{equation}
We recall that the time parameter $t'$ is to be chosen such that  the averaged horizon areas agree.   Clearly for   very  long times the $I_{BH}  \rightarrow I_{EBH}$ and all the original information is washed out.  The time derivative of the $I(t)$ gives the rate information is degenerated into noise.  

On the other hand, the entropy radiated by a black hole is 
\begin{equation}
\dot{S}= \frac{1}{125\hbar^3}  T_{BH} ^3 A \, ,
\end{equation}
 where $T_{BH} $ is Bekenstein's-Hawking temperature and $A$, the event horizon area . We can express the entropy flow in terms of the area eigenvalue [ eq.(\ref{qarea})]:
 \begin{equation}
\dot{S_n}= \  =\frac{1}{ 16 \sqrt{\kappa \hbar}}\int e^{-\lambda n} \lambda^{-1/2} d\lambda \, .
\end{equation}
As we discussed, he black hole is not in an eigenstate area, it is in a superposition of quantum states with probability given by  eq.(\ref{eternal}). Thus the information carried by the radiation  is actually:
 \begin{equation}
 \langle\dot{I}_{BH}(t)\rangle = \frac{1}{ 16 \sqrt{\kappa \hbar}}\int \lambda^{-1/2} d\lambda \sum_n e^{-\lambda n}\Pi_n(t) \, .
\end{equation}

With the help of the binomial distribution probability eq.(\ref{nbinomial}) and the convolution eq. (\ref{convolution}) we can express the information flow as
\begin{equation}
\langle\dot{I}_{BH}(t)\rangle   =\frac{(1-e^{-\alpha})^t}{ 16 \sqrt{\kappa \hbar}} \int_0^\infty \frac{\lambda^{-1/2} d\lambda}{(1-e^{(\lambda -\alpha)})^t} \sum_n e^{-\alpha n} \Pi_n(0) \, .
\end{equation}
This information embodies both signal and the noise, the later is represented by the "information" flow of an eternal black hole. A similar calculation yields
\begin{equation}
\langle\dot{I}_{EBH}(t')\rangle   =\frac{(1-e^{-\alpha})^{t'}}{ 16 \sqrt{\kappa \hbar}} \int_0^\infty \frac{\lambda^{-1/2} d\lambda}{(1-e^{(\lambda -\alpha)})^{t'}} 
\end{equation}
the time parameter $t'$ has to be chosen such that eternal black hole average mass is identical to black-hole itself.
The net information (the signal) flow at time $t$ is the difference $ \langle\dot{I}_{BH}(t)\rangle  - \langle\dot{I}_{EBH}(t)\rangle  $. The important question to be asked  is whether the rate useful information is degraded  is compensated by the rate  net information flows  . Should this be the case, the black hole conundrum might have found a solution. This is under investigation.
\section*{acknowledgment}
I am thankful to Nadav Sherb and Mikhail Zubkov for enlightening discussions.

 \end{document}